\documentclass[prd,twocolumn]{revtex4}
\usepackage{dcolumn}
\usepackage{array}
\usepackage{multirow}
\usepackage{booktabs}
\usepackage{graphicx}
\usepackage{amssymb}
\usepackage{bm}
\usepackage{hyperref}
\usepackage{epstopdf}
\usepackage{color}
\usepackage{mathrsfs}
\usepackage{amsmath,amssymb,amsthm}
\begin{document}
\title{Finslerian Universe May Reconcile Tensions between High and Low Redshift Probes}
\author{Deng Wang}
\affiliation{Theoretical Physics Division, Chern Institute of Mathematics, Nankai University,
Tianjin 300071, China}
\author{Xin-He Meng}
\affiliation{Department of Physics, Nankai University, Tianjin 300071, China}
\begin{abstract}
To reconcile the current tensions between high and low redshift observations, we perform the first constraints on the Finslerian cosmological models including the effective dark matter and dark energy components. We find that all the four Finslerian models could alleviate effectively the Hubble constant ($H_0$) tension and the amplitude of the root-mean-square density fluctuations ($\sigma_8$) tension between the Planck measurements and the local Universe observations at the 68$\%$ confidence level. The addition of a massless sterile neutrino and a varying total mass of active neutrinos to the base Finslerian two-parameter model, respectively, reduces the $H_0$ tension from $3.4\sigma$ to $1.9\sigma$ and alleviates the $\sigma_8$ tension better than the other three Finslerian models. Computing the Bayesian evidence, with respect to $\Lambda$CDM model, our analysis shows a \textit{weak} preference for the base Finslerian model and \textit{moderate} preferences for its three one-parameter extensions. Based on the model-independent Gaussian Processes, we propose a new linear relation which can describe the current redshift space distortions data very well. Using the most stringent constraints we can provide, we have also obtained the limits of typical model parameters for three one-parameter extensional models.

\end{abstract}
\maketitle
\section{Introduction}
In the past nearly two decades, a variety of cosmic probes such as supernova Type Ia (SNIa) \cite{1,2}, baryonic acoustic oscillations (BAO) \cite{3} and cosmic microwave background (CMB) anisotropies \cite{4,5} have confirmed the standard cosmological paradigm, namely the so-called inflationary $\Lambda$-cold-dark-matter ($\Lambda$CDM) model. Although recent measurements of CMB anisotropies by the Planck satellite \cite{6} have verified, once again, the high efficiency of the $\Lambda$CDM model in explaining the cosmological phenomena, there are still two main tensions between high and low redshift cosmic probes. On the one hand, the Hubble expansion rate derived indirectly from the latest Planck data \cite{7} under the assumption of the $\Lambda$CDM model is lower than the direct local measurement from Riess et al. over the $3\sigma$ confidence level (C.L.) by using improved SNIa calibration techniques \cite{8} (hereafter $H_0$ tension). On the other hand, the derived $\sigma_8$ parameter, the amplitude of the rms density fluctuations today in linear regime, from Planck data is, nonetheless, higher than the same quantity measured by several low redshift surveys including lensing, cluster counts and redshift space distortions (RSD) \cite{7,9,E} (hereafter $\sigma_8$ tension).

Up to date, it is still unclear that these tensions are originated from unknown systematic uncertainties in different methods used for measurements, or possibly small deviations from the $\Lambda$CDM model at all, indicating new physics needed ? In order to alleviate one or both of the above discrepancies, several solutions have been proposed, including dynamical dark energy \cite{10,11,12}, unified dark fluid \cite{13}, dark radiation \cite{14,15}, modified gravities \cite{16,17}, extended parameter space \cite{18,19}, cosmic voids \cite{20}, decaying dark matter \cite{21,22,WM}, ultralight axions \cite{23}, and so forth. Since the realistic nature of dark sector at the present time is still unknown, it is interesting and reasonable to give a possible explanation by exploring the connection between gravitation and new geometry. Modified Einstein's gravity may throw new light on the discrepancies.

\textit{Finsler geometry} \cite{24,25,26}, which involves Riemann geometry as its special case, opens naturally a new prospect to understand the current cosmological puzzles. This new geometry keeps the elegant properties of Riemann geometry, i.e., the isometric group is a Lie group on a Finslerian manifold, while it admits less Killing vectors than a Riemannian spacetime does. In general, there are at most $n(n-1)/2+1$ independent Killing vectors in a $n$ dimensional non-Riemannian Finslerian spacetime. Considering the simplest possible asymmetrical generalization of Riemannian metric, G. Randers \cite{27} proposed the well-known Randers space, a subclass of Finslerian space. In the framework of Randers space, a generalized Friedmann-Robertson-Walker (FRW) cosmological model based on Randers-Finsler geometry has been discussed \cite{28} and a modified dispersion relation of free particles has also been studied \cite{29}.

In history, the gravitational aspects of a Finslerian space were investigated for a long time \cite{30,31,32,33}. The gravitational field equations (GFEs) derived from a Riemannian osculating metric were exhibited in Ref. \cite{34}. For such a metric, the FRW-like cosmological scenario and the anisotropies of the Universe were also studied \cite{28,35}. Nonetheless, their GFEs are derived without considering the Bianchi identity and the general covariance principle of Einstein's gravity. It is interesting that, in Refs. \cite{36,c1,c2}, the authors have overcome these difficulties and derived the corresponding GFEs by constructing a Randers-Finsler space of approximate Berwald type (hereafter Berwald space), which is just an extension of a Riemannian space.

In this Letter, we focus on exploring whether the cosmic evolutional model derived from the Berwald space may reconcile the current $H_0$ and $\sigma_8$ tensions between high and low redshift observations. Our main result is that the Finslerian cosmological models considered in this analysis could alleviate effectively $H_0$ and $\sigma_8$ tensions.

\textit{Model.}--- For a Finsler-Berwald FRW Universe, the square of the effective dimensionless Hubble parameter $E(z)$ \cite{36} is expressed as
\begin{equation}
\begin{aligned}
E^2(z)= &\frac{9\alpha+\beta+12}{4(\alpha+1)(\beta+3)}\Omega_{m}(1+z)^{\frac{6(3\alpha+\beta+6)}{9\alpha+\beta+12}} \\
&+\frac{9(1-\omega)\alpha+(1+3\omega)\beta+12}{4(\alpha+1)(\beta+3)}\Omega_{de}                        \\
&\times (1+z)^{\frac{6[3(1-\omega)\alpha+(1+3\omega)\beta+6(1+\omega)]}{9(1-\omega)\alpha+(1+3\omega)\beta+12}},
\end{aligned}
\label{1}
\end{equation}
where $z$, $\omega$, $\Omega_{m}$, $\Omega_{de}$, $\alpha$ and $\beta$ denote the redshift, equation of state (EoS) of dark energy, effective matter and dark energy density parameters today, and effective time-component and space-component parameters, respectively. Here \textit{`` effective ''} means that the physical quantities are derived based on non-Riemannian Berwald space. Note that this model reduces to $\Lambda$CDM when $\omega=-1$ and $\alpha=\beta=0$. Since we are of much interest in studying the evolution of the late Universe, we ignore the contribution from the radiation component. Due to the fact that the spatial curvature of our Universe is very close to zero \cite{6}, for simplicity, we also ignore the contribution of the effective \textit{`` curvature ''} and just consider a flat Finsler-Berwald FRW Universe (see also Ref. \cite{36}).

To investigate how the Finslerian universe reconcile the global with local measurements, we take five cosmological models into account. The base $\Lambda$CDM model with the usual set of cosmological parameters: $\{ \Omega_bh^2, \Omega_ch^2, 100\theta_{MC}, \tau, \mathrm{ln}(10^{10}A_s), n_s \}$, where $\Omega_bh^2$ and $\Omega_ch^2$ are, respectively, the baryon and CDM densities today, $\theta_{MC}$ is the ratio between sound horizon and angular diameter distance at the decoupling epoch, $\tau$ is the Thomson scattering optical depth due to reionization, $\mathrm{ln}(10^{10}A_s)$ and $n_s$ are the amplitude and spectral index of primordial scalar perturbation power spectrum at the pivot scale $K_0=0.05$ Mpc$^{-1}$, respectively. Here $h\equiv H_0/100$ and $H_0$ is the Hubble constant. The simplest Finslerian cosmological model is a two-parameter extension of $\Lambda$CDM which corresponds to the case $\omega=-1$ in Eq. (\ref{1}). Hereafter we call it F$\Lambda$ model, which is the most important model considered in this analysis. In order to test overall the abilities of different Finslerian models to relieve tensions and verify the stabilities of our results with respect to (w.r.t.) assumptions on Finsler geometry, we also consider three one-parameter extensions to the F$\Lambda$ model: (i) The Finslerian $\omega$CDM (F$\omega$) model with constant EoS of dark energy; (ii) The Finslerian active neutrino (F$\nu$) model which allows the summed mass of three species of neutrinos $\Sigma m_\nu$ to vary with a degenerate hierarchy; (iii) The Finslerian sterile neutrino (Fs$\nu$) model allowing the effective number of relativistic species $N_{eff}$ to vary but $\Sigma m_\nu=0.06$ eV fixed.

\section{Analysis}
To perform constraints on the model parameters, we have modified carefully the publicly Markov Chain Monte Carlo code CosmoMC \cite{37,38} and Boltzmann code CAMB \cite{39}. Meanwhile, we choose flat priors to the above model parameters and marginalize the foreground nuisance parameters provided by Planck.  We use the temperature and polarization measurements from Planck (P) \cite{6}, RSD (R) data from the latest `` Gold-2017 '' compilation in Ref. \cite{40} and `` other '' (O) data sets such as SNIa data from the JLA compilation \cite{41}, BAO data \cite{42,43,44,45}, cosmic chronometer data from Ref. \cite{46}, HST data from Ref. \cite{8}, weak lensing data from CFHTLenS survey \cite{47} and CMB lensing data \cite{48}. For simplicity, we denote different data combinations as P, PR and PRO, respectively. A comprehensive analysis about the impacts of different data sets on the Finslerian scenarios are exhibited in a follow-up study \cite{49}.
\begin{table}[h!]
\renewcommand\arraystretch{1.3}

\caption{The constraints on the cosmological parameters of the F$\Lambda$ model at the 68$\%$ C.L. using different data combinations: P, PR and PRO.}
\label{t1}
\begin{tabular} { l c c c}

\hline
\hline
 Parameter & P & PR &  PRO\\
\hline
{$\Omega_b h^2   $} & $0.02123^{+0.00014}_{-0.00022}$ & $0.02149^{+0.00014}_{-0.00016}$ & $0.02240\pm 0.00013        $\\

{$\Omega_c h^2   $} & $0.1233^{+0.0019}_{-0.0015}$ & $0.11985\pm 0.00065        $ & $0.11696\pm 0.00084        $\\

{$100\theta_{MC} $} & $1.04016^{+0.00033}_{-0.00044}$ & $1.04065^{+0.00031}_{-0.00039}$ & $1.04112\pm 0.00027        $\\

{$\tau           $} & $0.0881^{+0.0045}_{-0.0034}$ & $0.0857\pm 0.0052          $ & $0.0811^{+0.0120}_{-0.0093} $\\

{$\alpha         $} & $0.0007^{+0.0032}_{-0.0023}$ & $0.00231^{+0.00120}_{-0.00096}$ & $0.0067^{+0.0073}_{-0.0090}$\\

{$\beta          $} & $0.0013^{+0.0020}_{-0.0036}$ & $0.0032^{+0.0028}_{-0.0039}$ & $0.0017\pm 0.0038         $\\

{${\rm{ln}}(10^{10} A_s)$} & $3.1096\pm 0.0064 $ & $3.1000\pm 0.0086          $ & $3.087^{+0.021}_{-0.017}   $\\

{$n_s            $} & $0.9510^{+0.0040}_{-0.0055}$ & $0.9600\pm 0.0031          $ & $0.9733^{+0.0040}_{-0.0035}$\\
\hline

$H_0                       $ & $65.04^{+0.62}_{-0.95} $ & $66.60^{+0.33}_{-0.36}     $      & $68.53\pm 0.41            $\\

$\sigma_8                  $ & $0.8480^{+0.0062}_{-0.0055}$  & $0.8350\pm 0.0049$  & $0.8236^{+0.0082}_{-0.0055}$                                                  \\
\hline
\hline
\end{tabular}
\end{table}
\begin{figure}
\centering
\includegraphics[width=5cm,height=3.8cm]{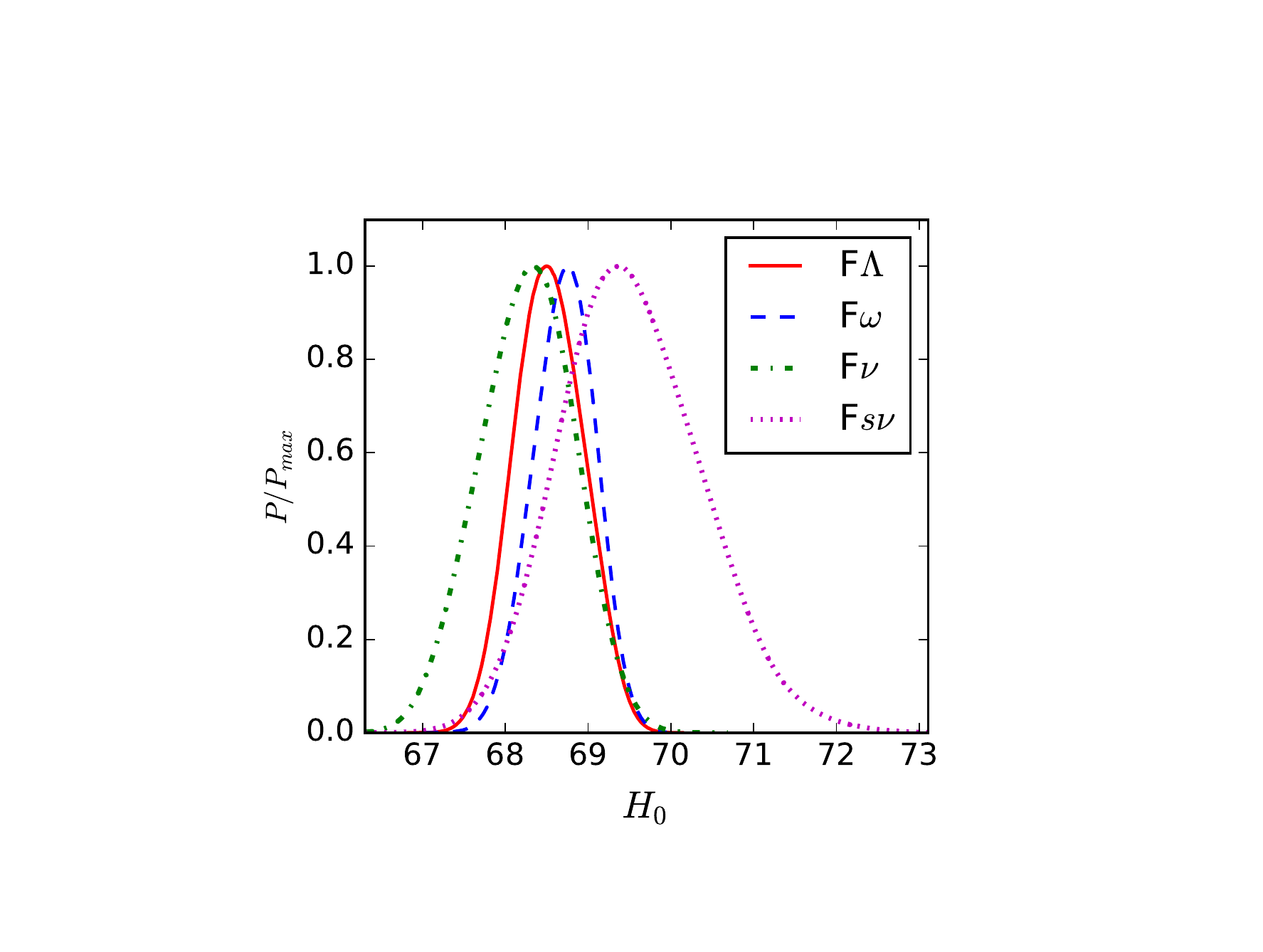}
\caption{The one-dimensional posterior probability distributions of $H_0$ values derived from the F$\Lambda$ (solid), F$\omega$ (dashed), F$\nu$ (dash-dotted) and F$s\nu$ (dotted) models using PRO data sets, respectively.}\label{f1}
\end{figure}

\begin{figure}
\centering
\includegraphics[width=2.5cm,height=2.5cm]{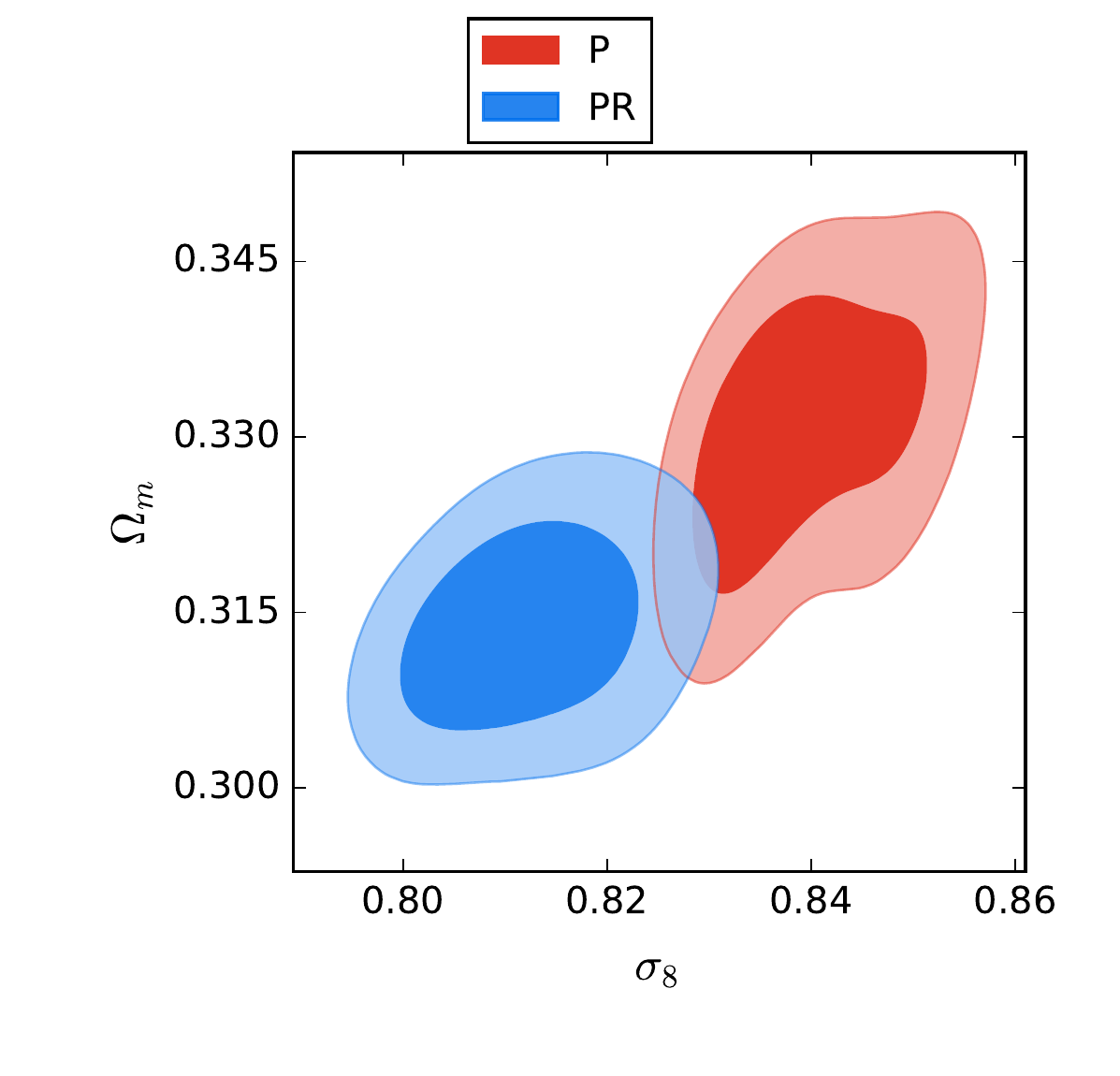}
\includegraphics[width=2.5cm,height=2.5cm]{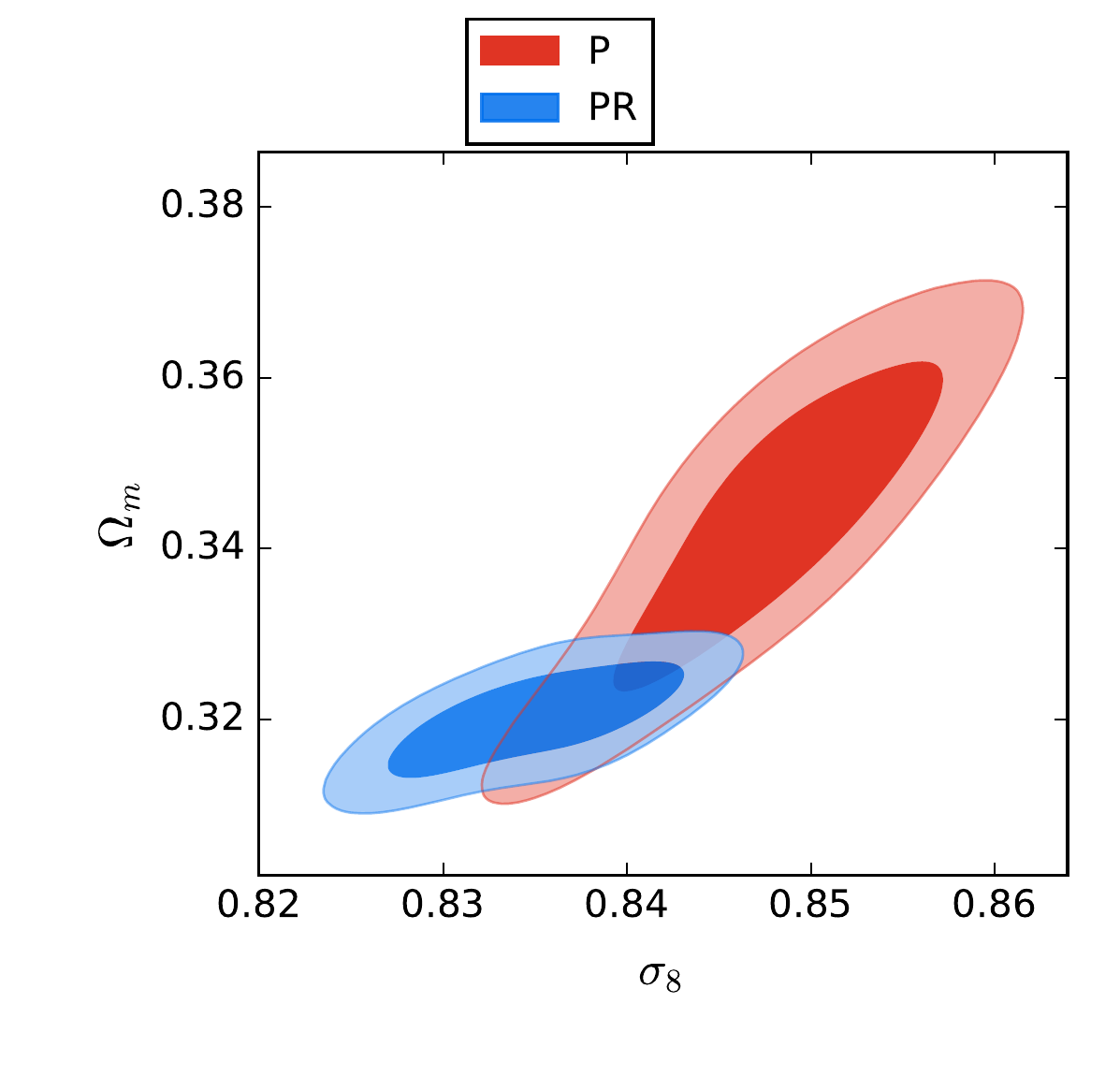}
\includegraphics[width=2.5cm,height=2.5cm]{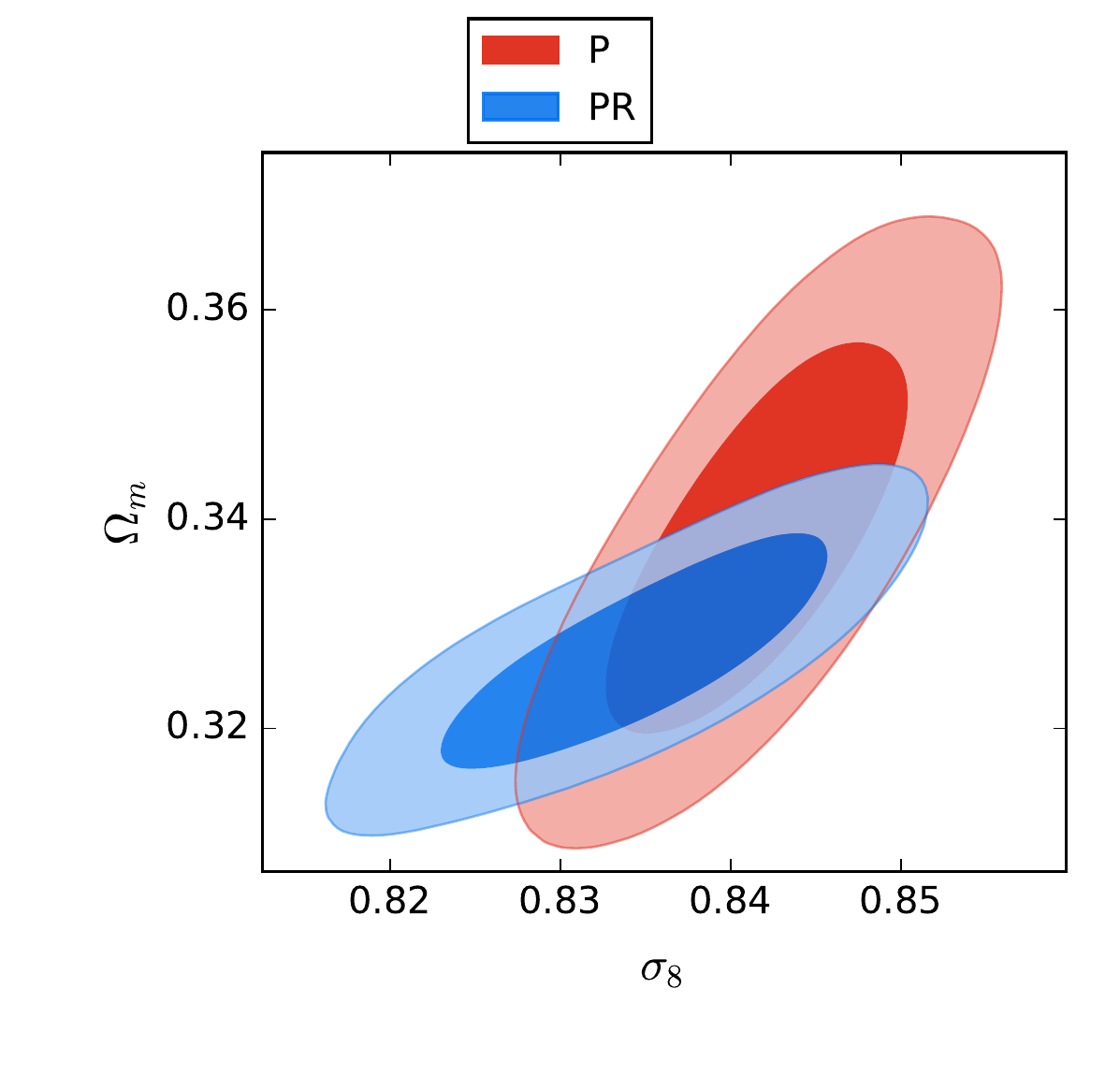}
\includegraphics[width=2.5cm,height=2.5cm]{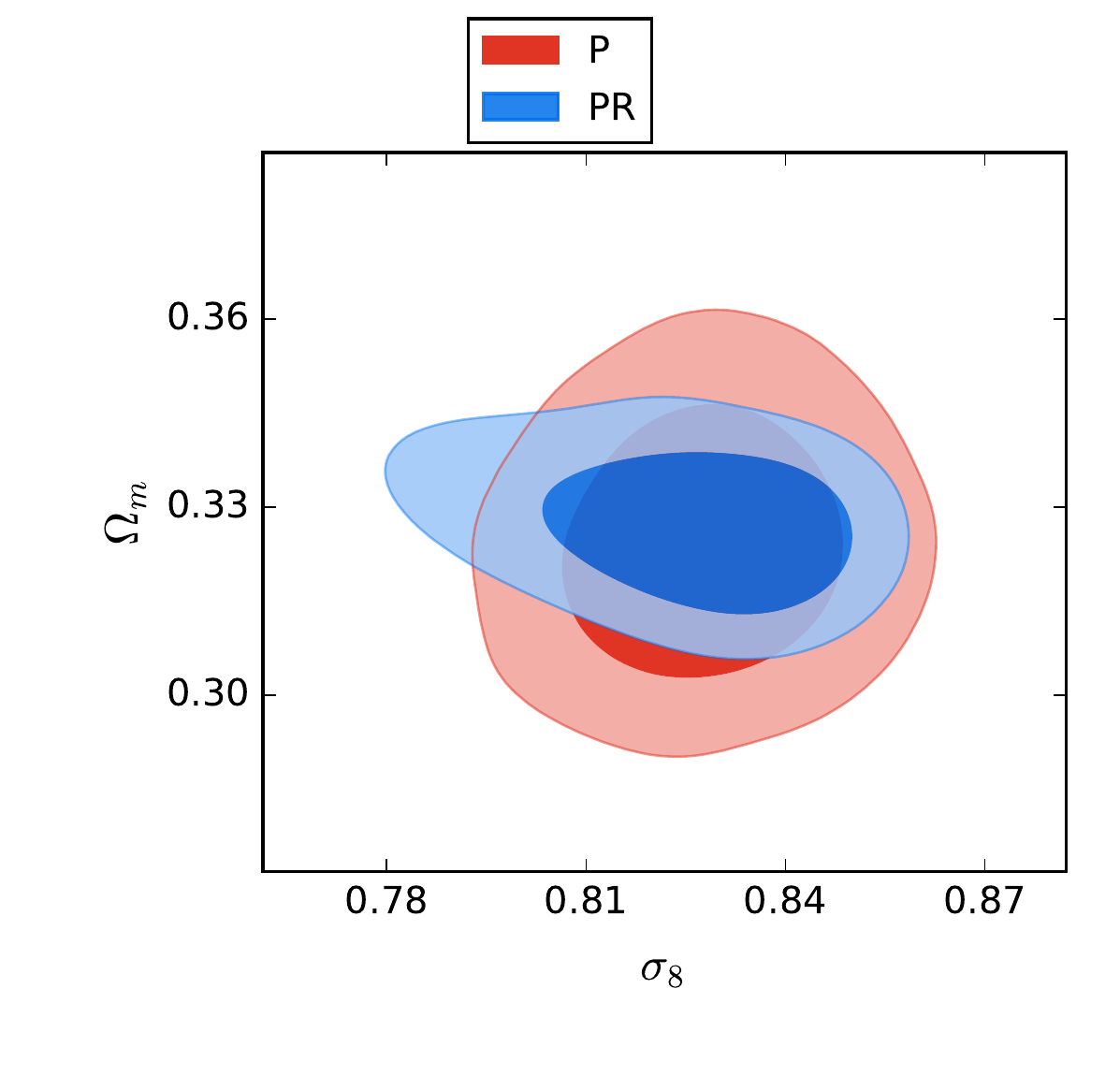}
\includegraphics[width=2.5cm,height=2.5cm]{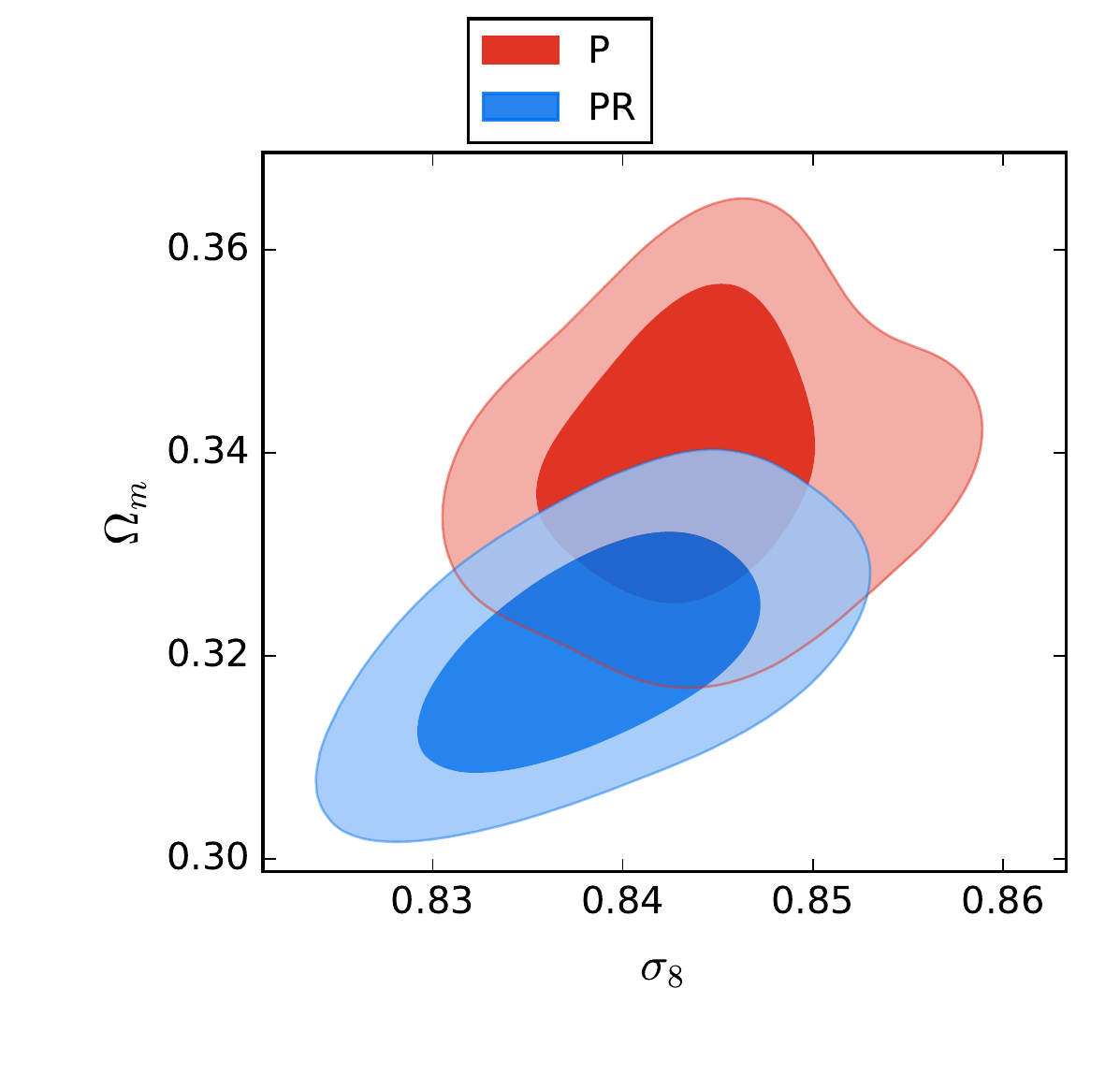}
\caption{From left to right, we show, respectively, the $\sigma_8-\Omega_{m}$ contours at the $68\%$ and $95\%$ C.L. from P (red) and PR (blue) data sets for the $\Lambda$CDM, F$\Lambda$, F$\omega$, F$\nu$ and F$s\nu$ models in turn.}\label{f2}
\end{figure}

\begin{figure}
\centering
\includegraphics[scale=0.4]{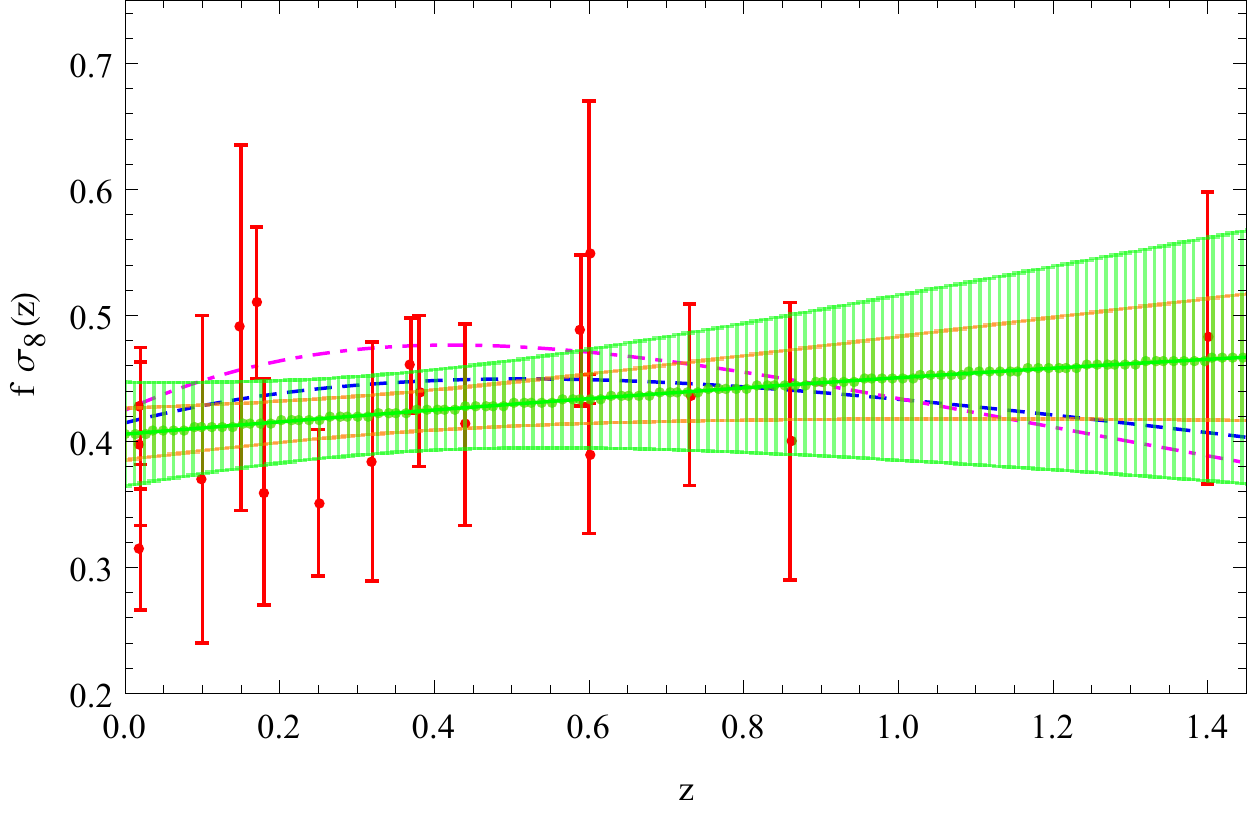}
\caption{The RSD data \cite{40} plotted against the theoretical predictions from the best-fit $\Lambda$CDM (dash-dotted) and F$\Lambda$ (dashed) models using PRO data sets. The green (solid) line represents the underlying true model producing these data. The orange and green error bars are GP reconstructions with 68$\%$ and 95$\%$ C.L., respectively.}\label{f3}
\end{figure}

\begin{table}[h!]
\renewcommand\arraystretch{1.2}
\caption{The minimum $\chi^2$ values, Bayesian evidence and strength of evidence for different cosmologies using the combined constraints PRO.}
\begin{tabular}{l c c c}
\hline
\hline
      Model        & $\chi^2_{min}$    & $\ln B_{ij}$  & \quad Strength of evidence               \\
\hline
$\Lambda$CDM       &13741.664                    &---            &---                      \\
F$\Lambda$         &13735.052                    &1.363          &weak            \\
F$\omega$          &13731.528                    &2.721          &moderate              \\
F$\nu$             &13731.116                    &2.883          &moderate                  \\
F$s\nu$            &13727.840                    &4.205          &moderate                     \\
\hline
\hline
\end{tabular}
\label{t2}
\end{table}

The constraining results of the F$\Lambda$ model using different data combinations are presented in Tab. \ref{t1}. Interestingly, both parameters $\alpha$ and $\beta$ are consistent with zero at the 68$\%$ C.L. and both prefer slightly the positive values. This means that a small deviation from the $\Lambda$CDM model has been realized in the framework of Finsler geometry, and the evolution of the late Finslerian Universe remains dominated by dark energy. We also find that the $H_0$ value derived from PRO can alleviate the $H_0$ tension between Planck ($H_0=66.93\pm0.62$ \cite{7}) and local measurements ($H_0=73.24\pm1.74$ \cite{8}) more effectively than those derived from P and PR. Furthermore, we show the posterior probability distributions of $H_0$ values derived from four Finslerian models in Fig. \ref{f1}. Opening an extra parameter $\omega$, $\Sigma m_\nu$ or $N_{eff}$, we find, as expected, a mild relaxation on $H_0$ bounds at the 68$\%$ C.L.. Particularly, in the F$s\nu$ model, this tension has been reduced from $3.4\sigma$ to $1.9\sigma$. Meanwhile, we find that, using PRO, the constraint on the EoS of dark energy $\omega=-1.00017^{+0.00090}_{-0.00069}$ in the F$\omega$ model is not only well consistent with $\omega=-1.006\pm0.045$ in the $\omega$CDM model from Planck data, but also is more stringent than Planck's prediction by one order of magnitude. (We also investigate later the effects of two one-parameter Finslerian models on $H_0$.)

We also notice that, to some extent, the distinct Finslerian scenarios can alleviate the current $\sigma_8$ tension. From Fig. \ref{f2}, one can find that, for four Finslerian models, the parameter pair ($\sigma_8, \Omega_{m}$) constrained by P are all compatible with that constrained by PR at the $68\%$ C.L.. In particular, the F$\nu$ model can alleviate the tension better than other three Fisnlerian models. Attractively, three two-parameter extensions of F$\Lambda$ model perform a little better than itself in terms of $\sigma_8$ tension. In Tab. \ref{t1}, one can also find that the joint constraint from PRO for the F$\Lambda$ model supports a lower $\sigma_8$ value than other two data combinations P and PR.

RSD are very important probes of large scale structure (LSS) providing bias-free data of $f\sigma_8(z)$. In light of significantly overlapped data today, we use the relatively robust and independent $f\sigma_8(z)$ data  compiled in Ref. \cite{40} from different surveys to compare our constrained F$\Lambda$ model with $\Lambda$CDM. In Fig. \ref{f3}, we find that, interestingly, the evolutional behavior of the F$\Lambda$ model describes better the low-$z$ data and the unique high-$z$ data located at $z=1.40$ than $\Lambda$CDM, i.e., having lower values when $z<1.01$ and higher values $z>1.01$ (here $z=1.01$ is the cross point where two best-fit models have the same prediction). Since it seems that neither of these two models give an accurate depiction to current RSD data, to study the underlying mechanism phenomenologically generating RSD data, we employ the model-independent Gaussian Processes (GP) method to reconstruct $f\sigma_8$ versus $z$. Specifically, we use the publicly package GaPP (Gaussian processes in python) to implement reconstructions (see Refs. \cite{50,51} for details on GP). We find that the best-fit F$\Lambda$ model is consistent with the underlying true model form GP reconstructions at the 2$\sigma$ C.L. and that the best-fit $\Lambda$CDM model lies out of the 2$\sigma$ confidence region when $0.11\lesssim z\lesssim0.57$ (see Fig. \ref{f3}). This means that the F$\Lambda$ model may be a little better than $\Lambda$CDM in terms of only RSD data. More interestingly, we find a linear relation (hereafter LR), $f\sigma_8(z)=0.042z+0.406$, that can describe current RSD data very well.

\begin{figure}
\centering
\includegraphics[width=3.1cm,height=3.1cm]{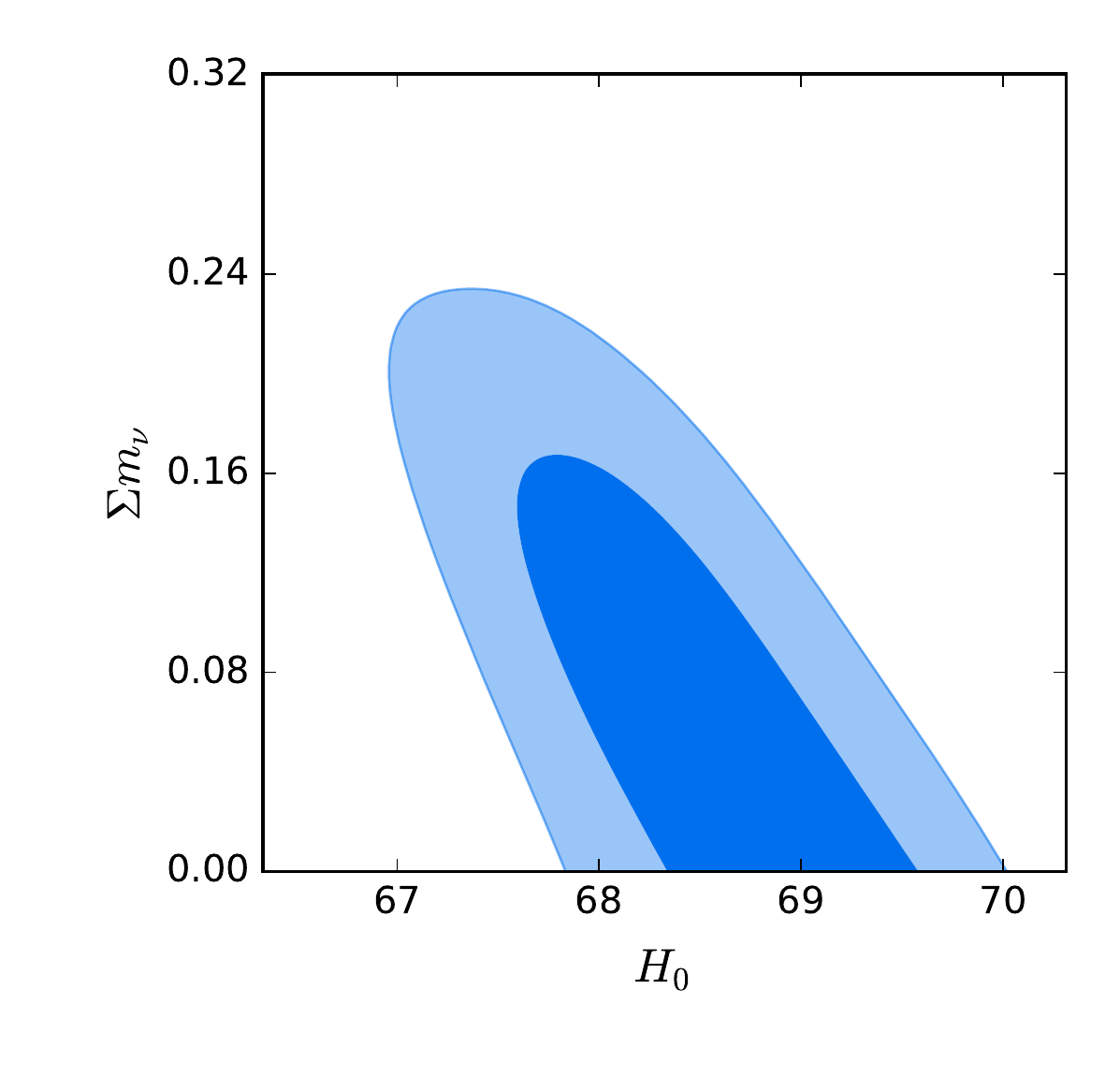}
\includegraphics[width=3.1cm,height=3.05cm]{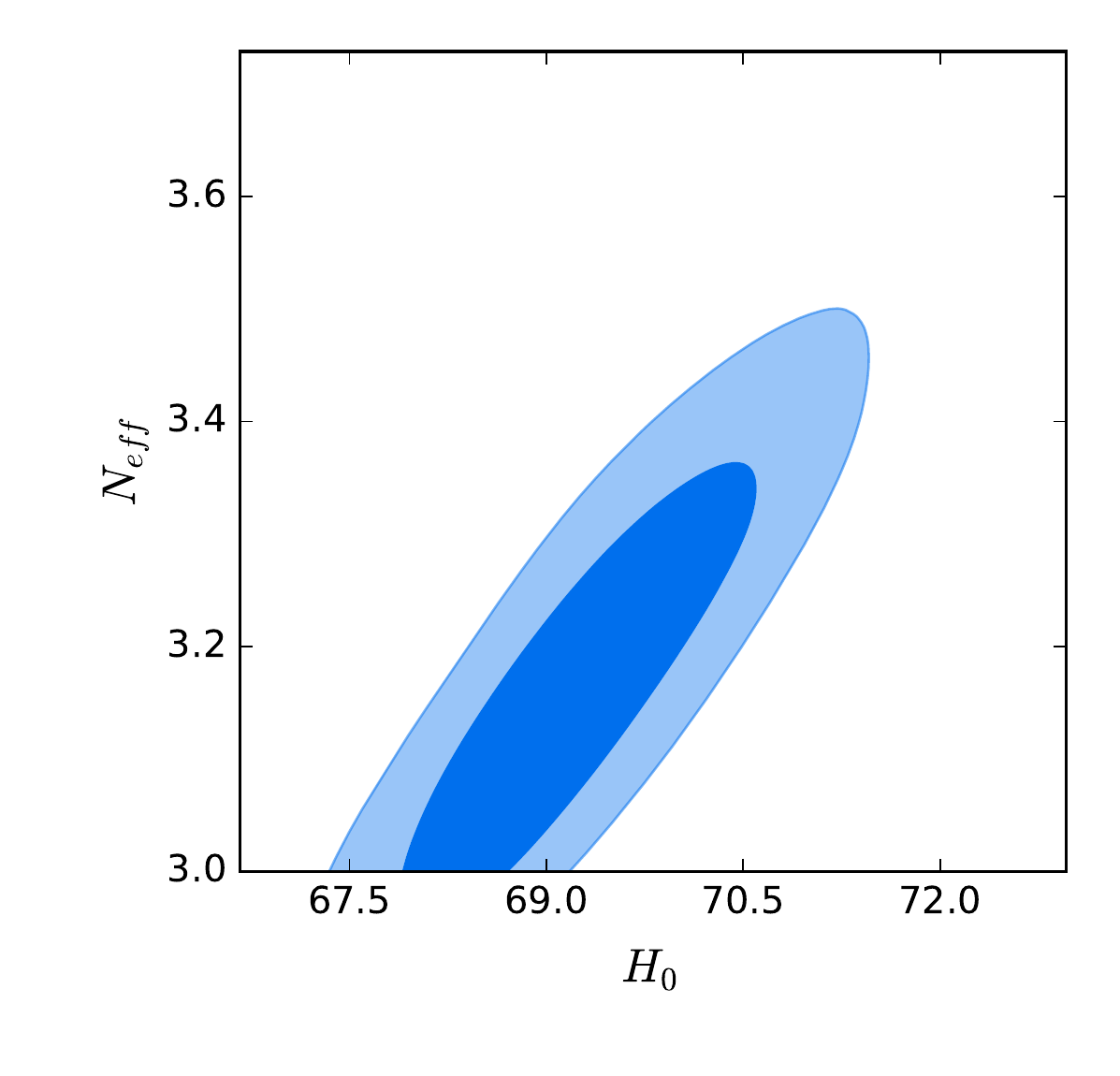}
\caption{The two-dimensional posterior distribution contours in the $\Sigma m_\nu-H_0$ plane for the F$\nu$ model and in the $N_{eff}-H_0$ plane for the F$s\nu$ model, respectively.}\label{f4}
\end{figure}

In order to explore the observational viability of different models, choosing $\Lambda$CDM as the reference model, we compute the corresponding Bayesian evidence of four Finslerian models, $\varepsilon_i$, and Bayes factor, $B_{ij}=\varepsilon_i/\varepsilon_j$, where $\varepsilon_j$ denotes the evidence of the reference model. Following Ref. \cite{52}, we adopt a revised and more conservative version of the so-called Jeffreys' scale, i.e., $\ln B_{ij} = 0 - 1$, $\ln B_{ij} = 1 - 2.5$, $\ln B_{ij} = 2.5 - 5$ and $\ln B_{ij} > 5$ indicate an \textit{inconclusive}, \textit{weak}, \textit{moderate} and \textit{strong} preference of the model $i$ relative to the reference model $j$. Note that for an experiment which leads to $\ln B_{ij} < 0$, it means the reference model is preferred by data.

Tab. \ref{t2} shows the minimum $\chi^2$ values, logarithmic Bayes factors and strength of evidence, respectively, for various Finslerian theoretical models by using PRO. Clearly, we have obtained a better fit to current data w.r.t. $\Lambda$CDM. Meanwhile, it is interesting that, w.r.t. $\Lambda$CDM,  the analysis shows a \textit{weak} preference for the F$\Lambda$ model and \textit{moderate} preferences for its three one-parameter extensions F$\omega$, F$\nu$ and F$s\nu$. We also note that, although four Finslerian scenarios give a better description for PRO data sets than $\Lambda$CDM, such a result is obtained at the cost of increasing significantly the value of reionization optical depth $\tau$. For example, the value $\tau=0.0811^{+0.0120}_{-0.0093}$ in the F$\Lambda$ model is 2.02$\sigma$ higher than the previous estimation $\tau=0.055\pm0.009$ derived with Planck HFI data \cite{7}.

It is also attractive to investigate the fitting performances of two one-parameter extensions to $\Lambda$CDM by setting either the effective space-component parameter $\beta=0$ (hereafter F$\alpha$ model) or time-component parameter $\alpha=0$ (F$\beta$). Utilizing the PRO data sets, we find that the analysis also shows a \textit{weak} preference for the F$\alpha$ ($\ln B_{ij} = 1.126$) or F$\beta$ ($\ln B_{ij} = 1.312$) model, and that the $H_0$ tension can be reduced from 3.4$\sigma$ to 2.7$\sigma$ (F$\alpha$) and 2.66$\sigma$ (F$\beta$), respectively.

Recently, based on the high-precision Planck data, in order to explore the radiation components of the early Universe, researchers have tried to place tighter constraints than before on $\Sigma m_\nu$ and $N_{eff}$ \cite{14,15,53}, which modifies the damping tail of the CMB temperature angular power spectrum. We are also interested in addressing these issues in the framework of Finsler geometry. Using the most stringent constraints from PRO, we obtain $\Sigma m_\nu < 0.186$ eV in the F$\nu$ model and $N_{eff} <3.40$ in the F$s\nu$ model at the 95$\%$ C.L., respectively (see Fig. \ref{f4}). Furthermore, to exhibit how $\Sigma m_\nu$ and $N_{eff}$ affects the constraints on $H_0$, we plot the two-dimensional posterior distribution contours in the $\Sigma m_\nu-H_0$ plane for the F$\nu$ model and in the $N_{eff}-H_0$ plane for the F$s\nu$ one. We find that $H_0$ is anti-correlated and positively correlated with $\Sigma m_\nu$ and $N_{eff}$, respectively.
Note that our constraint on the total mass of active neutrinos $\Sigma m_\nu < 0.186$ eV is larger than the bound 0.1 eV with a degenerate hierarchy measured by Planck data \cite{6}. This means that the F$\nu$ model is still a competitive candidate of dark energy. Meanwhile, as Planck measurements \cite{6}, the F$s\nu$ model excludes a fully thermalized neutrino ($\Delta N_{eff}=N_{eff}-3.046\approx1$) over $3\sigma$ C.L. using PRO.

\section{Discussions}
Our explorations on using the Finsler geometry to reconcile tensions among various cosmic probes are just in the beginning stage. It is noteworthy that, in this analysis, we do not utilize the Finslerian models to address the internal inconsistencies (e.g., the so-called $\tau$ and $A_{lens}$ tension  \cite{7}, where $A_{lens}$ denotes the amplitude of lensing power relative to the physical value) existing in Planck data \cite{6,7}. Based on the current RSD data, the new phenomenological LR could give some clear hints of the underlying model, which dominates the evolution of the late Universe. If one expects to alleviate easily the $\sigma_8$ tension by constructing a reasonable cosmological model, a model which can provide the evolution of $f\sigma_8(z)$ against $z$ being very close to the new LR may be a good candidate. The questions are whether one can find this kind of underlying models. If yes, what physical and mathematical conditions need they satisfy ? Furthermore, we expect that future high-quality RSD data can provide a good test of our new LR. In addition, recently, it has been suggested that, to a large extent, the properties of dark energy could influence the cosmological weighing and mass ordering of neutrinos. We also make attempts to investigate the neutrino sector using the Finslerian scenarios. It is worth noting that the massive sterile neutrinos either thermally or Dodelson-Widrow distributed \cite{54} in light of Finsler geometry are also needed to be explored. Moreover, we have only considered the linear growth of LSS and not investigated the nonlinear evolution of the effective Finslerian dark matter and dark energy. The remaining issues are addressed in a forthcoming work \cite{49}.

\section{Conclusions}
We have primarily tested the abilities of Finslerian cosmological models in alleviating the current $H_0$ and $\sigma_8$ tensions between high and low redshift observations. Using the data combination PRO, we find that all the four Finslerian models could alleviate effectively the $H_0$ tension at the $68\%$ C.L., and that, especially, this tension has been reduced from $3.4\sigma$ to $1.9\sigma$ in the F$s\nu$ model. Similarly, the $\sigma_8$ tension can also be effectively alleviated by these four models at the $68\%$ C.L.. Particularly, the F$\nu$ model can alleviate the tension better than other three models. Interestingly, we also find that the parameters $\alpha$ and $\beta$ of both F$\Lambda$ model are in good agreement with zero at the 68$\%$ C.L. and both prefer slightly the positive values. Computing the Bayesian evidence, we find that the Finslerian models give a better fit to current data. Specifically, w.r.t. $\Lambda$CDM, our analysis shows a \textit{weak} preference for the F$\Lambda$ model and \textit{moderate} preferences for its three one-parameter extensions F$\omega$, F$\nu$ and F$s\nu$. Moreover, w.r.t. $\Lambda$CDM, current data also exhibits a weak \textit{weak} preference for two one-parameter extensions to the F$\Lambda$ model, F$\alpha$ and F$\beta$ models, which could also alleviate the $H_0$ tension at the $68\%$ C.L.. In terms of current RSD data, based on the GP reconstructions, we also propose a new LR, which characterizes the RSD data very well. Attractively, using PRO data sets, the constraint on the EoS of dark energy $\omega=-1.00017^{+0.00090}_{-0.00069}$ in the F$\omega$ model is not only well compatible with $\omega=-1.006\pm0.045$ in the $\omega$CDM model from Planck data, but also is more stringent than Planck's prediction by one order of magnitude. We also give $95\%$ limits of the total mass of active neutrinos $\Sigma m_\nu < 0.186$ eV with a degenerate hierarchy in the F$\nu$ model and the effective number of relativistic species $N_{eff}<3.40$ in the F$s\nu$ model, respectively.

\section{acknowledgements}
We thank B. Ratra and S. D. Odintsov for helpful discussions about cosmology. Deng Wang thanks Jing-Ling Chen, F. Melia, Yang-Jie Yan, Qi-Xiang Zou and Yuan Sun for useful communications. This study is supported in part by the National Science Foundation of China.

\end{document}